# Nationwide Hourly Population Estimating at the Neighborhood Scale in the United States Using Stable-Attendance Anchor Calibration

Huan Ning[1, 2*], Zhenlong Li[2*], Manzhu Yu[2], Xiao Huang[1], Shiyan Zhang[2,3], Shan Qiao[4]

[1] Department of Environmental Sciences, Emory University, Atlanta, GA, United States
[2] Geoinformation and Big Data Research Laboratory, Department of Geography, Pennsylvania State University, University Park, PA, United States
[3] Department of Geography and Environmental Studies, New Mexico State University, Las Cruces, NM, United States
[4] South Carolina SmartState Center for Healthcare Quality, University of South Carolina, Columbia, SC, United States

Correspondence: zhenlong@psu.edu, huan.ning@emory.edu

**Abstract**: Traditional population datasets are largely static and therefore unable to capture the strong temporal dynamics of human presence driven by daily mobility. Recent smartphone-based mobility data offer unprecedented spatiotemporal coverage, yet translating these opportunistic observations into accurate population estimates remains challenging due to incomplete sensing, spatially heterogeneous device penetration, and unstable observation processes. We propose a Stable-Attendance Anchor Calibration (SAAC) framework to reconstruct hourly population presence at the Census block group level across the United States. SAAC formulates population estimation as a balance-based population accounting problem, combining residential population with time-varying inbound and outbound mobility inferred from device-event observations. To address observation bias and identifiability limitations, the framework leverages locations with highly regular attendance as calibration anchors, using high schools in this study. These anchors enable estimation of observation scaling factors that correct for under-recorded mobility events. By integrating anchor-based calibration with an explicit sampling model, SAAC enables consistent conversion from observed device events to population presence at fine temporal resolution. The inferred population patterns are consistent with established empirical findings in prior mobility and urban population studies. SAAC provides a generalizable framework for transforming large-scale, biased digital trace data into interpretable dynamic population products, with implications for urban science, public health, and human mobility research. The hourly population estimates can be accessed at: https://gladcolor.github.io/hourly_population.

**Keywords**: Dynamic population estimation; Human mobility; Mobile phone data; Sampling bias correction; Anchor-based calibration; Hourly population; Digital demography

## 1   Introduction

Population data are fundamental to studies of human society, including urban planning, public health, social sciences, resource distribution, and emergency management. Governments and administrators invest substantial effort in obtaining the population data in a timely manner to understand how many people live and work in jurisdictions, as well as their demographic characteristics (Baffour et al. 2013). Knowledge of population distribution and its temporal trends supports infrastructure planning, public service provision, and forecasting of future societal needs.

Traditional data collection methods, such as decennial censuses conducted every decade in the United States (U.S.) and China, provide detailed and comprehensive demographic information but often suffer from substantial temporal lag, with data aggregation and release often delayed by 2 or 3 years (Bureau 2022). To mitigate this lag, intermediate surveys such as the American Community Survey 1-year and 5-year estimates (ACS) are widely used. These products contain population size, distribution, and demographic characteristics at the Census block group (CBG) level and are commonly applied in policy analysis and scientific research (Hattis 2020). The spatial resolution of Census data can be very detailed since it surveys every household, but such data will not be released due to privacy concerns. The commonly used fine-grained product level is at the CBG level, with each unit containing 1000 – 3000 residents, aggregated from the minimum areal unit of blocks.

Modern societies, however, are inherently dynamic, with population distributions varying continuously across space and time. Long-term changes may arise from economic development, environmental conditions, housing construction, or migration, occurring over years or decades at metropolitan or neighborhood scales. At much shorter temporal scales, daily commuting represents the most prominent form of population redistribution. Residents of "bedroom communities" or "sleeping cities" routinely travel to employment centers or central business districts during weekday daytime hours. Such daily mobility patterns are ubiquitous in urban environments and directly shape the design and operation of transportation systems, including public transit networks and road infrastructure. Beyond transportation, urban facilities such as parks, grocery stores, and healthcare services are profoundly influenced by hourly and daily fluctuations in population presence. Static residential population estimates alone are insufficient to support planning and management of these facilities.

Accurate estimation of population surges during large events is also critical for public safety. Failures to anticipate crowd size can have catastrophic consequences. For example, a crowd crush during Halloween festivities in Itaewon, South Korea in October 2022, resulted in more than 150 fatalities (Mao 2023; Sharma et al. 2023), while a New Year's Eve stampede in Shanghai in 2014 caused dozens of deaths and injuries (Zhou et al. 2018). Religious gatherings and festivals can similarly generate dramatic short-term population increases (Gayathri et al. 2017). Beyond crowd safety, dynamic population estimates are increasingly important for hazard exposure assessment, such as air pollution or heat exposure, where hourly population distributions aligned with real-time environmental observations provide more accurate risk estimates than residential population alone (Gariazzo et al. 2016; Picornell et al. 2019; Yu et al. 2023; 2024).

Dynamic population patterns arise from diverse forms of human mobility, including commuting by workers and students (McKenzie 2013; Fowler 2024; Laughlin et al. 2015), tourism (Tian et al. 2023), hospital visits, shopping, and nightlife activities. Researchers have explored various methods using different data sources to depict the dynamic population (Panczak et al. 2020) and city rhythms (Güller and Varol 2024; Gökçe Kılıç and Terzi 2025). Early efforts to quantify such dynamics combined multiple administrative and survey data sources. For example, Moss and Qing (2012) synthesized census data, commuter statistics, student enrollments, hospital records, and visitor counts to estimate weekday population dynamics in Manhattan, revealing a substantial discrepancy between residential population (1.62 million), census-based daytime population (3.07 million), and estimated weekday presence (3.94 million). This work highlighted the magnitude of population redistribution that remains invisible in static demographic datasets.

With the widespread adoption of mobile phones, researchers have increasingly relied on mobile phone-based data as proxies for population dynamics. Call detail records (CDRs) and mobile signaling data (MSD), which record interactions between mobile devices and cellular towers during phone calls, text messaging, and data usage, have been widely employed due to the high penetration of mobile phones, with smartphone adoption in the United States estimated at 97% by 2022. Mobility-derived population measures are particularly valuable in regions where reliable census data are unavailable or outdated (Deville et al. 2014). Deville et al. (2014) analyzed billions of CDRs from Portugal and France to estimate real-time population density by correlating observed phone activity with census population distributions. Similarly, Liu et al. (2018) and Li et al. (2019) reconstructed individual trajectories from sparse CDR observations to derive hourly population estimates in Chinese cities. Such approaches, however, remain constrained by cellular tower coverage, uneven sampling, and carrier-specific biases. Using similar datasets, Xu et al. (2015) analyzed aggregate mobility patterns and activity spaces in Shenzhen, China, while Zaragozí et al. (2021) examined seasonal tourism flows in Costa Daurada, Spain.

More recently, large-scale mobility datasets derived from smartphone application pings have become available, including products such as Advan (formerly SafeGraph) (Advan Research 2022a) and Veraset (veraset.com). These datasets provide anonymized, aggregated visitation records for points of interest (POIs), along with inferred home locations of devices. They have been widely applied to studies of disease transmission (Ning et al. 2023; Li, Qiao, et al. 2024; Chang et al. 2021), environmental exposure (Yu et al. 2024; Akinboyewa et al. 2025), and tourism (Kupfer et al. 2021). At the same time, substantial work has examined the representativeness, biases, and limitations of these datasets (Li et al. 2021; Li, Ning, et al. 2024; Jiang et al. 2025). A common challenge across applications is the need to relate observed device-event counts to the underlying population, often addressed through normalization by an assumed device sampling rate (Kang et al. 2020).

Several recent studies have focused on improving spatial allocation of dynamic populations by incorporating land use, building footprints, and activity context. Wei et al. (2023) used mobile phone signaling data to generate spatiotemporal human activity heatmaps in Xining, China, while Bergroth et al. (2022) estimated 24-hour population distributions by allocating population to building floor areas using dasymetric interpolation. Related work has highlighted differences in population activity across land cover. Global population products such as LandScan provide daytime and nighttime population estimates using land cover and ancillary data (Weber et al. 2022). Although these efforts have substantially improved spatial resolution, most remain limited in their ability to capture continuous temporal dynamics at fine spatial scales.

Despite extensive progress in high-resolution population mapping (WorldPop 2024; Huang et al. 2021; Weber et al. 2022), relatively few studies have achieved large-area population estimation with both high spatial resolution (e.g., CBG level) and high temporal resolution (e.g., hourly). Dynamic population estimation remains constrained by data availability, heterogeneous data providers, and the lack of transparent, generalizable calibration methods. While social media data have been used to estimate daily population variation at coarse spatial (Martín et al. 2021; Huang et al. 2022), universal approaches for fine-grained, hourly population reconstruction are still rare.

Additionally, among studies on dynamic populations using diverse datasets, the structural properties of data collection and observation modeling are under investigation.

To address these gaps, we propose a Stable-Attendance Anchor Calibration (SAAC) framework that uses smartphone-based mobility data (SafeGraph Patterns; now Advan Patterns) to reconstruct hourly population presence for each Census block group across the United States. This study produces one of the first nationwide hourly population datasets at neighborhood scale, enabling analysis of dynamic population patterns across hourly, providing a valuable complement to static population products such as ACS and LandScan.

## 2 Results

### 2.1 Overview of Stable-Attendance Anchor Calibration (SAAC) framework

We develop a Stable-Attendance Anchor Calibration (SAAC) framework to estimate hourly population presence at the Census block group (CBG) level by integrating static residential population data with passively observed human mobility signals. The framework formulates population estimation as a balance-based accounting problem, in which the population present in each CBG at a given hour is determined by the residential population adjusted by time-varying inbound and outbound mobility flows. While residential population is directly obtained from census data, mobility-derived components are inferred from event-based observations collected through smartphone-derived mobility datasets.

Figure 1 provides an overview of the Stable-Attendance Anchor Calibration (SAAC) framework, illustrating how biased and incomplete mobility observations are transformed into hourly population presence through anchor-based calibration and population balance constraints.

SAAC explicitly models the mobility observation process as an incomplete and biased sampling mechanism, in which only a fraction of population presence generates observable device events. To correct for under-recording and spatially heterogeneous device penetration, the framework introduces *observation scaling factors* that map observed device events to population counts. Because these scaling factors are not directly identifiable at fine spatial and temporal resolutions, SAAC leverages locations with highly regular attendance patterns as calibration anchors. In this study, U.S. high schools serve as stable-attendance anchors, enabling empirical estimation of observation scaling factors under conditions of minimal temporal variability. Stable-attendance anchors provide an identifiability gain by introducing place–time contexts in which the latent population size is approximately constant, allowing the observation rate to be empirically estimated.

The calibrated scaling factors are subsequently transferred to the broader mobility dataset to estimate hourly inbound population presence across all CBGs. Hourly outbound population is inferred using a constrained reconstruction approach that enforces consistency between inbound and outbound flows at each time step. Together, these components form a coherent and scalable framework for reconstructing hourly population dynamics from biased and incomplete mobility observations. The following sections describe each component of the SAAC framework in detail.

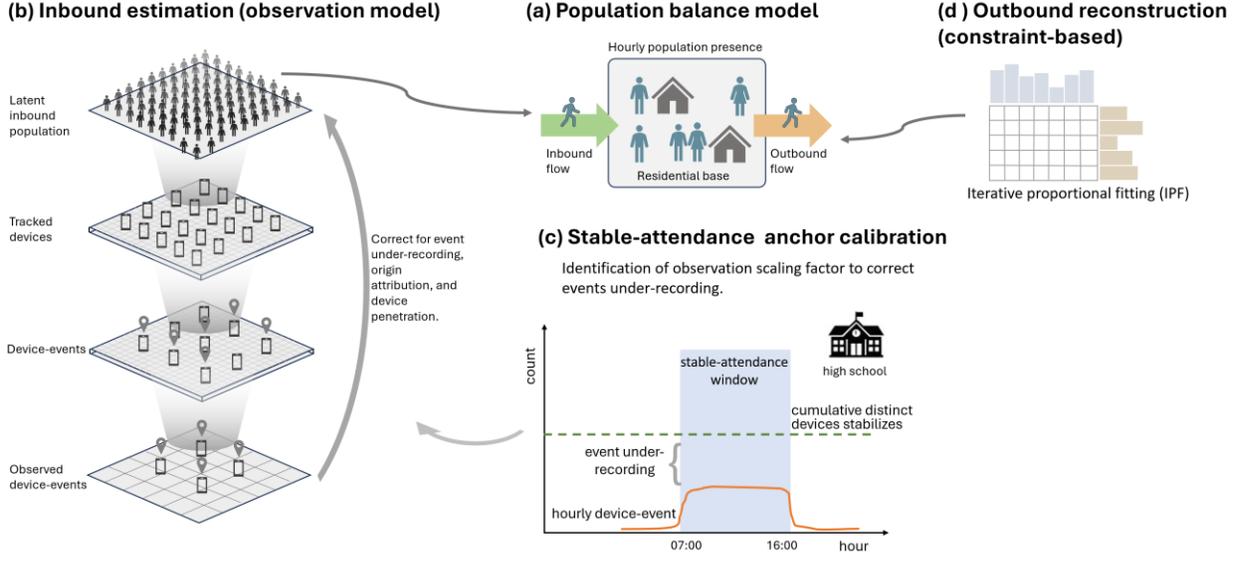

Figure 1 Overview of the Stable-Attendance Anchor Calibration (SAAC) framework. (a) Population balance model defining hourly population presence as the residential base adjusted by time-varying inbound and outbound flows. (b) Estimation of inbound population presence from passively observed device events through correction for event under-recording, origin attribution, and heterogeneous device penetration. (c) Calibration of observation scaling factors using stable-attendance anchors (e.g., high schools), where distinct device counts stabilize under minimal temporal variability. (d) Reconstruction of hourly outbound flows via iterative proportional fitting to ensure consistency with inferred inbound presence and aggregated origin–destination information. Panels are read in the order (a)–(d).

## 2.2 Population balance model

We estimate hourly population presence at the CBG level using a population balance framework that combines residential population with time-varying inbound and outbound mobility. Let $p_c^t$ denote the population present in CBG $c$ at hour $t$. The hourly population presence is expressed as:

$$p_c^t = N_c + In_c^t - Out_c^t \tag{1}$$

where $N_c$ is the residential population of CBG $c$ obtained from the ACS, $In_c^t$ denotes the inbound population presence entering CBG $c$ at hour $t$, and $Out_c^t$ denotes the outbound population leaving CBG $c$ at the same hour. A full notation table is provided in Supplementary Information.

Equation (1) enforces a population balance constraint at the hourly scale: the population present in each CBG is determined by its residential baseline, augmented by inbound mobility and reduced by outbound mobility. This balance equation forms the conceptual backbone of the SAAC framework. While the residential population $N_c$ is directly obtained from ACS data, both $In_c^t$ and $Out_c^t$ must be inferred from passively collected mobility data under incomplete and heterogeneous observation. The remainder of the methodology focuses on estimating these mobility-derived components through explicit observation modeling and anchor-based calibration.

## 2.3 Stable-attendance anchor calibration (SAAC)

A broad class of existing statistical methods has addressed problems analogous to those posed by mobile device–based population estimation, namely inferring latent population sizes from incomplete and biased observations. In ecology ("Mark–Recapture" 2009), epidemiology (Chao et al. 2001), and survey statistics (Couso and Sánchez 2011), capture–recapture techniques and multiplier methods (Wang et al. 2024) are routinely used to estimate hidden or partially observed populations by exploiting repeated observations under stable conditions. Similarly, calibration estimators and inverse probability weighting adjust observed samples to match known population totals by estimating an observation or inclusion probability. In these approaches, the central challenge is identifiability: observed counts reflect the product of true population size and an unknown observation rate, and additional structure, such as repeated sampling, known anchors, or auxiliary information, is required to disentangle the two. These methods share a common mathematical foundation with mobile population estimation, in which digital traces represent a biased and incomplete sample of the underlying population rather than a census.

In the context of mobility data, related ideas have been implicitly adopted through national- or regional-level normalization strategies, where observed device counts are scaled to known census populations under assumptions of spatial and temporal homogeneity (Kang et al. 2020; Advan Research 2025a; 2025b). However, such approaches typically rely on fixed or externally imposed scaling factors and do not explicitly model the observation process or its variability across time and space. In contrast, anchor-based calibration frameworks estimate observation rates empirically by leveraging locations with stable, known activity levels. The SAAC framework builds on this statistical lineage by introducing a transparent and data-driven mechanism to identify observation rates directly from mobility data, using institutions with predictable attendance patterns as internal calibration anchors. By embedding classical calibration principles into a spatiotemporally explicit observation model, SAAC extends these established statistical ideas to the problem of large-scale, high-resolution dynamic population reconstruction.

### 2.3.1 Observation model and calibration procedure

Many vendors provide mobility datasets that record digital traces generated by a subset of the population through mobile devices. We model this process as a two-stage mechanism involving event generation and partial observation. In the statistical formulation, we use the term *unit* to denote an abstract event-generating entity; in the empirical implementation, each *unit* corresponds to a single tracked mobile device. At any given hour $t$, a latent set of event-generating units, i.e., individuals carrying mobile devices and present in a CBG, may generate location events. Let $u$ denote the number of such units present during a stable attendance window. Importantly, $u$ is fixed but unobserved. Due to incomplete device participation, intermittent location reporting, and platform-specific recording rules, only a subset of generated events is observed. Let $e_t$ denote the number of observed events in hour $t$. The observed data therefore represent a partial observation of the underlying event-generating process, rather than a complete census.

We define the observation rate as the ratio between observed events and latent units,

$$observation\_rate_t = \frac{e_t}{u}$$

which captures the combined effects of device penetration and event under-recording. In practice, both $u$ and the observation rate are unknown, and only $e_t$ is directly observed.

The partial observability of mobility data leads to a fundamental identifiability problem. From a single observation window, the observed event count $e_t$ does not uniquely identify either the number of event-generating units $u$ or the observation rate. Multiple combinations of latent *unit* counts and observation rates can generate the same observed event count. As a result, population presence and mobility flows cannot be directly recovered without additional structure or assumptions.

This identifiability challenge is intrinsic to opportunistic digital trace data and cannot be resolved by increasing data volume alone. Instead, it requires external information or constraints that link observed events to the underlying population.

To address this identifiability challenge, SAAC introduces stable-attendance anchors: locations and time windows in which the underlying population presence is approximately constant and predictable. Formally, a stable-attendance anchor is defined as a place–time context in which (i) the number of event-generating units remains stable over a known window, and (ii) repeated observations can be accumulated without substantial changes in the latent population. High schools during weekday school hours are a canonical example of such anchors. Within a stable-attendance window, the cumulative number of distinct observed units increases over time and eventually stabilizes. This stabilization provides information about the underlying number of event-generating units, enabling estimation of the observation rate.

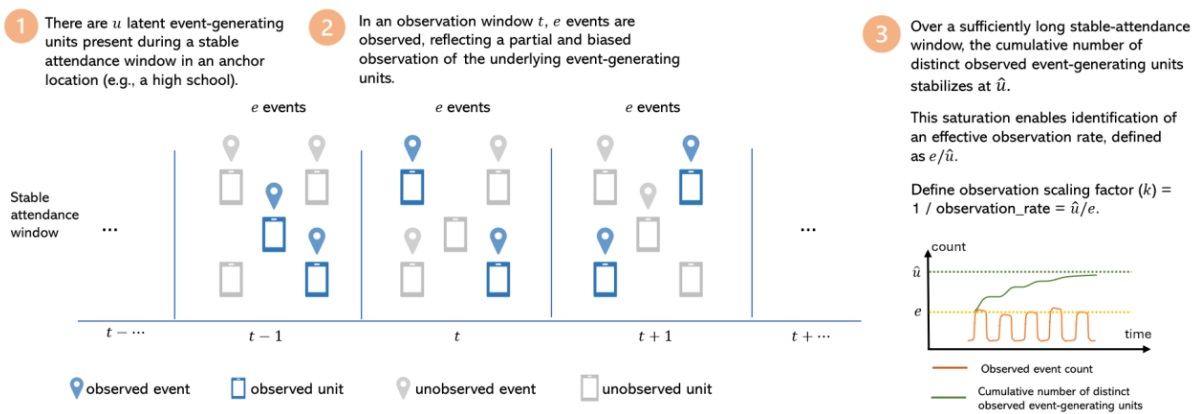

Figure 2 Conceptual illustration of the observation model and calibration procedure. During a stable attendance window at an anchor location, a fixed number of latent event-generating units are present. Only a partial and biased subset of their events is observed within each time window. Over a sufficiently long stable-attendance period, the cumulative number of distinct observed units stabilizes, enabling identification of an effective observation rate and its inverse sssssscaling factor, which are used to calibrate mobility observations to population presence.

Figure 2 is a conceptual illustration of SAAC, showing its major steps. SAAC leverages the presence of dual observation processes in mobility datasets: event-level observations and unit-level observations (e.g., distinct devices observed over a longer aggregation window).

By accumulating observations across multiple time windows within a stable-attendance anchor, the cumulative number of distinct observed units converges to an estimate $\hat{u}$ of the true number of event-generating units. Once $\hat{u}$ is obtained, the observation rate is defined as

$$\widehat{\text{observation\_rate}} = \frac{e}{\hat{u}}$$

where $e$ denotes a typical observed event count within the stable window. The corresponding observation scaling factor is defined as

$$k = \frac{1}{\widehat{\text{observation\_rate}}} = \frac{\hat{u}}{e}$$

This scaling factor provides a principled bridge between observed mobility events and latent population presence. The observation scaling factor estimated from stable-attendance anchors is assumed to be transferable across nearby spatial units and temporally proximate periods. In Figure 2, conceptually, there are 5 ($\hat{u}$) devices in an anchor location, and 2 ($e$) events observed are observed; thus, $k = \hat{u}/e = 5/2 = 2.5$, indicating that an observed event may reflect 2.5 latent events.

In practice, SAAC aggregates calibration estimates to a county–month level to ensure robustness and mitigate local noise. Once calibrated, the scaling factor is applied to event-based mobility observations to infer inbound and outbound population flows at the CBG–hour level, thereby enabling reconstruction of hourly population presence across the study region.

### 2.3.2 High schools as stable-attendance calibration anchors

Section 2.3.1 introduces the Stable-Attendance Anchor Calibration (SAAC) framework and formalizes the observation model underlying event-to-unit scaling. In this section, we describe how SAAC is operationalized using U.S. high schools as stable-attendance anchors and illustrate the procedure with a concrete example.

High schools provide repeated observation windows with temporally stable attendance during weekday school hours. Using Weekly Patterns, we extract two complementary observations for each identified high school point of interest: hourly visit counts, which we treat as observed events, and weekly distinct device counts, which represent observed event-generating units. Weekday school hours, defined as 7:00 AM to 4:59 PM local time, are used as the stable-attendance anchor window.

Figure 3 illustrates the SAAC implementation for a representative high school. The figure displays all hourly visit counts across 52 weeks in 2022, organized by month. During routine school hours on school days (red dots), hourly event counts are highly stable across days and weeks, yielding a well-defined typical hourly event level. An increase in the fall semester (September–December) is also visible. During these stable periods, weekly distinct device counts remain proportional to the typical hourly event level, resulting in stable estimates of the observation scaling factor.

In contrast, weeks containing after-school hours (5:00 PM–9:59 PM) or weekend gatherings (blue and green dots), such as concerts, sporting events, or election polling, exhibit elevated weekly device counts relative to school-hour events. These weeks violate the stable-attendance assumption and are therefore excluded from calibration (indicated by grey dashed lines), for example, the week of September 5–9.

These patterns indicate that Weekly Patterns data capture both routine school-hour mobility and episodic non-routine disturbances. Stable-attendance anchors therefore provide a diagnostic baseline for assessing data consistency and sensitivity. The observed proportionality between school-hour events and weekly distinct devices supports the use of hourly event counts for identifying observation scaling factors used to extrapolate population presence.

For "clean" weeks without substantial visitation disturbances (green dashed lines, e.g., Figure 3, February 7–11 and February 14–18), we treat them as *anchor* weeks. A typical school-hour visit count is computed and used as the observed event count $e$ in the SAAC framework. The corresponding weekly distinct device count represents the cumulative number of distinct observed event-generating *units* over the same stable-attendance window. As described in Section 2.3.1, saturation of the cumulative number of observed units across repeated anchor windows enables estimation of the effective observation rate and, equivalently, the observation scaling factor $k$.

To maximize the applicability of SAAC across schools, we adopted a set of inclusion rules for identifying anchor weeks. Empirically, we observe that weekly distinct device counts stabilize after approximately four stable-attendance windows, corresponding to four school days. For example, the week of January 17–21 exhibits stable device counts relative to January 10–14, whereas weeks with only three school days yield substantially lower device counts (e.g., January 3–7 compared with January 10–14). Based on these observations, we require anchor weeks to contain at least four school days. In addition, weeks with after-school or weekend gatherings are retained if their non-school-hour event peaks remain below the typical school-hour event level. Additional selection rules and diagnostics are provided in the Supplementary Information.

Using this procedure, we used approximately 28,000 high schools and successfully derived observation scaling factors for about 16,000 schools spanning roughly 2,500 counties. For counties or months without sufficient anchor weeks to support direct calibration, observation scaling factors were imputed using values from the corresponding state or from temporally adjacent months. Nationally, the observation scaling factor $k$ is about 2.7.

Across the national set of high school anchors, this procedure yields a distribution of observation scaling factors that characterize the relationship between observed device *events* and latent event-generating *units* under stable-attendance conditions. These anchor-derived scaling factors are subsequently aggregated to the county–month level and incorporated into the CBG inbound population reconstruction stage, as described in Section 5.2.

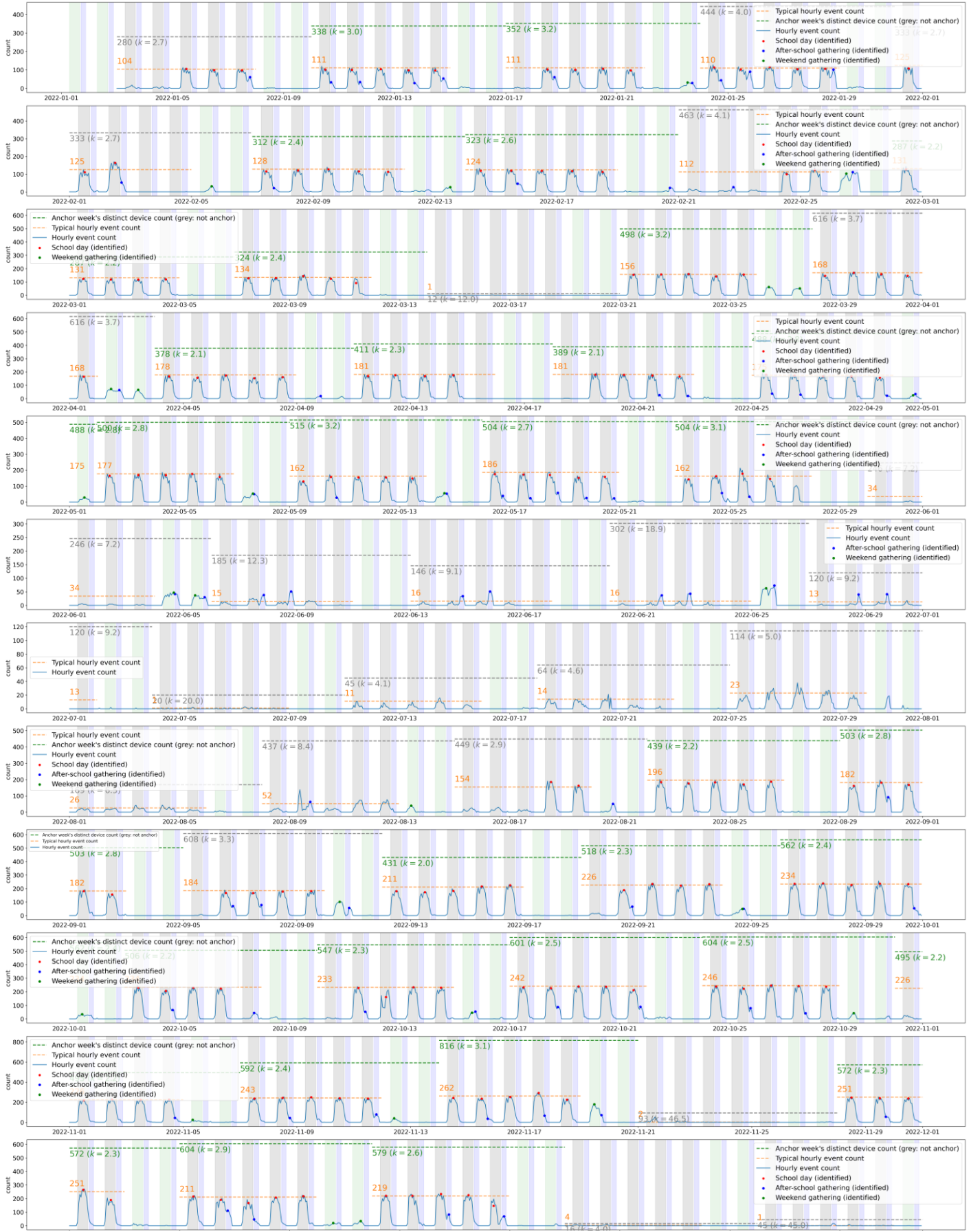

Figure 3 High school visit–device dynamics as stable-attendance anchor diagnostics. This figure plots the hourly visits of a high school. Grey spans: weekday school hours (7:00AM – 4:59PM); light green spans: weekend school hours; light blue spans: after-school hours (5:00PM – 9:59PM). School hour visits during school days (red dots) are highly stable across weeks, while after-school hours and weekends show low visitation with occasional spikes during non-routine events (blue and green dots). Weekly device counts reflect broader participation and increase disproportionately during event weeks. These patterns illustrate the suitability of high schools as stable-attendance anchors for diagnosing observation consistency and calibrating scaling factors in the SAAC framework.

## 2.4 Case Study: Reconstructed Hourly Population in Manhattan

Using the Stable Attendance Anchor Calibration framework, we reconstructed hourly population presence for all Census block groups (about 220,000) for the United States in 2022 from passively observed mobility data. Figure 4 presents the resulting population dynamics of Manhattan in January 2022 as a representative case.

Figure 4(a) shows Census block group level hourly population maps for a representative weekday, Wednesday, January 19, 2022. Clear diurnal transitions are observed across the city. During nighttime hours from 00:00 to 05:00, population presence is spatially concentrated in residential areas. From early morning onward, population presence increases rapidly, with pronounced buildup between 05:00 and 09:00. Midday hours from 10:00 to 15:00 exhibit stabilized population presence across major employment corridors, followed by a gradual decline during the evening from 16:00 to 23:00. A consistent color scale across hours highlights systematic spatial redistribution rather than local fluctuations.

In addition to the citywide diurnal pattern, major commuting hubs exhibit a distinctive temporal signature. As shown in Figure 4(a) and referenced in Figure 4(b), population presence at major commuter entry points to Manhattan, including the Penn Station Terminal, Grand Central Terminal, and the Brooklyn Bridge corridor, increases earlier in the morning than in surrounding areas and remains elevated later into the evening. These locations are consistently among the first areas to show population buildup during the morning commuting period and among the last to decline during the evening hours. This "early onset and late decay" pattern indicates that the reconstructed hourly population captures systematic commuting dynamics rather than transient or location-specific fluctuations.

A contrasting pattern is observed in Central Park, a major non-residential urban destination that attracts large numbers of local residents and visitors from around the world. Figure 4(b) and (c) show that both the ACS 2019 population and the LandScan population products assign negligible or zero population to this area, reflecting residential or land use-based assumptions embedded in these datasets. In contrast, the mobility-based reconstruction identifies a persistent daytime population presence within Central Park, with elevated levels emerging during morning hours and remaining throughout the day. Independent reports estimate approximately 115,000 visitors per day to Central Park (Alex Van Buren 2016; Greensward Group 2023; Central Park Conservancy 2022). The reconstructed population exhibits peak hourly presence of approximately 35,000 individuals, with about 15,000 during other daytime hours. Given that hourly population estimates represent the number of individuals present within each hour and may include transient pass through movement along adjacent roadways, these magnitudes are broadly consistent with reported visitation levels under differing measurement definitions.

Figure 4(d) plots the aggregated hourly population of Manhattan over January 2022. The reconstructed time series exhibits pronounced weekday diurnal cycles, with daytime peaks consistently exceeding the ACS 2019 residential baseline and similar to the LandScan (Weber et al. 2022) daytime estimates and the weekday daytime range reported by Moss and Qing (2012). During nighttime hours, the reconstructed population converges above the ACS baseline and approximates the nighttime estimates of Moss and Qing (2012); in contrast, LandScan nighttime estimates are below the ACS baseline. Weekend periods show attenuated daytime peaks and reduced temporal variability relative to weekdays. Across the full month, the relative ordering among weekday daytime, weekend daytime, and nighttime population levels remains consistent with prior empirical estimates. A marked population decline is observed on Saturday, January 29, corresponding to the documented impact of a major winter storm affecting the New York City region.

Together, the contrasting behaviors observed at commuting hubs and non-residential urban destinations illustrate that the reconstructed hourly population captures both structured commuting flows and activity-driven population presence that are not represented in residential or land use-based population datasets.

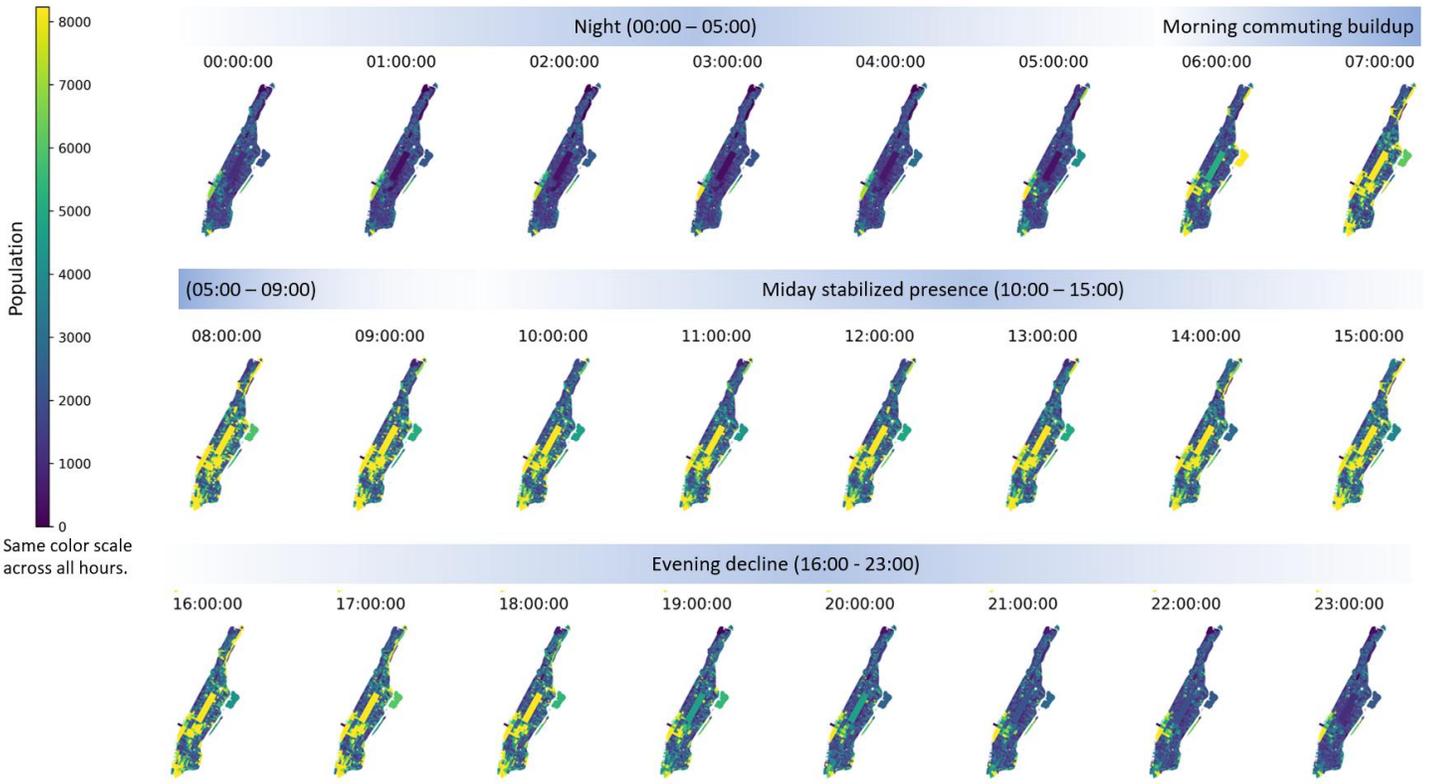
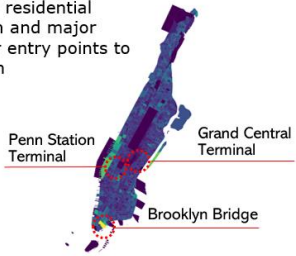
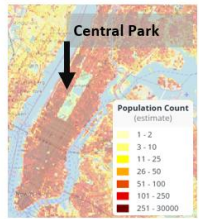
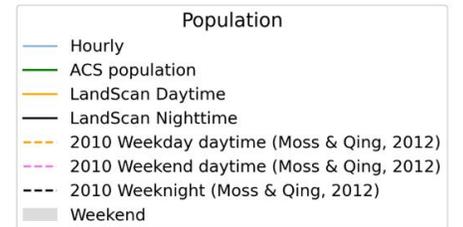
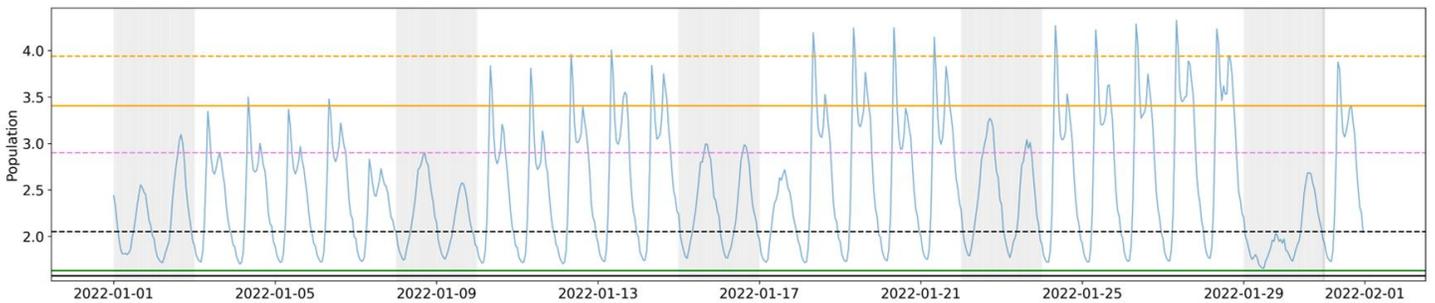

Figure 4 Hourly population dynamics of Manhattan reconstructed using SAAC. (a) Census block group level hourly population presence in Manhattan on January 19, 2022, reconstructed from smartphone mobility observations. A consistent color scale highlights nighttime residential concentration, morning commuting buildup, midday stabilization, and evening decline. (b) ACS residential population with major commuter entry points shown for reference. (c) LandScan daytime population estimates with popular Central Park excluded. (d) Aggregated hourly population for January 2022 compared with ACS and LandScan benchmarks.

# 3 Discussion

This study reframes high-frequency population estimation as a calibration-constrained inverse inference problem under partial observability, rather than a direct scaling or interpolation exercise. Smartphone-derived mobility observations represent an opportunistic, systematically biased sample of an underlying latent population process, shaped by heterogeneous device ownership, platform-specific reporting rules, and temporally varying user behavior. Without explicit constraints, observed device-event counts alone are insufficient to uniquely identify true population presence. The SAAC framework addresses this fundamental identifiability challenge by introducing empirically grounded constraints derived from stable-attendance contexts, thereby enabling principled reconstruction of hourly population dynamics.

## 3.1 Anchor-based identifiability and statistical grounding

At its core, SAAC adapts multiplier and capture–recapture logic from epidemiology and ecology (Chao 2001; Chao et al. 2001; Couso and Sánchez 2011), to the domain of digital mobility traces. In these classical settings, repeated observations under stable conditions provide the structural information necessary to disentangle population size from observation probability. Analogously, SAAC leverages high schools during weekday instructional hours as natural calibration anchors, where attendance is temporally regular, and the latent population can be treated as approximately constant across repeated observation windows. The stabilization of cumulative distinct device counts within these windows provides empirical leverage to estimate observation scaling factors that correct for event under-recording and incomplete device participation.

By explicitly modeling the observation process rather than treating it as a nuisance, SAAC resolves a limitation common to many mobility-based population products: the implicit assumption of spatially and temporally homogeneous device sampling. The resulting calibration is internal to the mobility data itself and does not require external enrollment records or assumed attendance sizes. This feature is particularly important at the national scale, where external administrative data are often incomplete, inconsistent, or unavailable.

## 3.2 Empirical coherence of reconstructed population dynamics

The reconstructed hourly population surfaces exhibit coherent and interpretable spatiotemporal patterns at the Census block group level across the United States. Diurnal cycles, weekday–weekend contrasts, and episodic disruptions emerge consistently and align with well-established empirical expectations. At aggregate scales, daytime population peaks exceed residential baselines in employment centers, while nighttime populations converge toward ACS residential estimates—patterns that closely mirror prior empirical estimates for Manhattan (Moss and Qing 2012) and national ambient population products such as LandScan (Weber et al. 2022).

Importantly, the agreement with independent benchmarks is not imposed by design but emerges as a consequence of anchor-calibrated inference. Sensitivity analyses show that implausible values of $k$ (e.g., $k = 1$) produce population estimates that are inconsistent with independent benchmarks of Moss and Qing (2012), reinforcing the need for anchor-based calibration.

### 3.3 Beyond ratio-based normalization

Many existing approaches to mobility-derived population estimation rely on ratio-based normalization, scaling observed device counts to known population totals under assumptions of proportionality. While operationally convenient, such approaches may be sensitive to diurnal variation in device usage, particularly during nighttime hours when devices may be inactive or sparsely reporting, and to aggregation across time zones, which conflates distinct local observation regimes.

SAAC generalizes and extends these approaches by replacing global homogeneity assumptions with locally and temporally grounded calibration. Ratio-based methods can be understood as a limiting special case of observation scaling under strong assumptions; SAAC relaxes these assumptions by empirically estimating observation rates where identifiability is achievable and transferring them cautiously across space and time. This distinction is critical for applications requiring fine temporal resolution, such as environmental exposure assessment or emergency response.

### 3.4 Scientific and practical contributions

Beyond the specific population dataset produced in this study, SAAC contributes a general methodological framework for transforming biased digital traces into interpretable population measures. A key contribution is that SAAC does not require direct knowledge of anchor population sizes or explicit enrollment data. Instead, it exploits structural properties of repeated observation under stable attendance to infer observation rates empirically. This makes the framework relatively easy to implement and transferable across datasets that contain both event-level and unit-level observations. For mobility data providers, anchor-based calibration offers a transparent diagnostic tool for evaluating observation consistency and temporal stability, even when anchor POIs are excluded from downstream products (e.g., the exclusion of schools in newer Weekly Patterns Plus releases). For researchers, SAAC provides an external benchmark against which simulation-based mobility models, e.g., agent-based or synthetic population systems, can be evaluated and stress-tested.

Conceptually, this work demonstrates the feasibility of embedding classical statistical inference principles into large-scale digital mobility analysis. By framing population estimation as an inverse problem constrained by stable-attendance anchors, SAAC bridges the gap between raw device events and population science, offering a replicable and theoretically grounded alternative to ad hoc normalization strategies.

### 3.5 Limitations and future research

SAAC relies on the assumption that device usage behavior at anchor locations is representative of surrounding populations. If smartphone usage at high schools differs systematically from other settings, derived scaling factors may be biased when transferred. In addition, the framework assumes relatively stable origin CBG distributions for observed devices within county–month units, whereas actual distributions may vary by time of day or day of week. While these assumptions are mitigated through aggregation and balance constraints, relaxing them remains an important direction for future work.

We found that approximately 2% of Census block groups produced "negative" hourly population estimates during the outbound estimation stage, necessitating further post-processing. These implausible values are likely caused by the combined effects of the iterative proportional fitting (IPF) procedure, inaccuracies in the assumed origin distributions of devices, and the lagged ACS data. While IPF ensures that reconstructed flows match marginal totals, it does not enforce behavioral realism or non-negativity constraints on intermediate values. If the underlying origin distribution poorly reflects actual mobility patterns, especially in low-population or data-sparse areas, IPF may overestimate outbound flows and generate negative net population counts. Addressing this issue may require more updated ACS baseline data, robust initialization strategies, detailed device origin distribution, adaptive smoothing, or explicit constraints in the reconstruction pipeline.

Several avenues for future research emerge from this work. Methodologically, SAAC could be extended by incorporating additional anchor types, such as workplaces, transit hubs, or residential nighttime anchors, enabling stratified or multi-anchor calibration. Comparing derived population estimates with independent enrollment or administrative records, where available, would further validate anchor assumptions. From an application perspective, hourly population estimates derived through SAAC can support environmental exposure assessment, disaster risk analysis, and public health studies that require temporally resolved population denominators. Integrating these estimates into geospatial foundation models and large-scale urban simulations represents a promising direction for advancing dynamic population science.

## 4 Conclusion

This study introduces Stable-Attendance Anchor Calibration (SAAC), an anchor-calibrated inverse inference framework for reconstructing fine-grained, hourly population dynamics from passively collected mobile phone data under partial observability. By leveraging locations with temporally stable attendance patterns, SAAC provides a principled mechanism to identify observation rates and bridge the gap between observed device events and latent population presence. The framework adapts long-established multiplier and capture–recapture concepts from epidemiology and ecology to the context of large-scale digital mobility data, offering a transparent and empirically grounded alternative to global normalization approaches.

Applied at the national scale, SAAC enables reconstruction of hourly population presence at the Census block group level across the U.S. The resulting population estimates exhibit coherent diurnal and weekday–weekend patterns that are broadly consistent with independent population benchmarks and prior empirical studies. These results demonstrate that anchor-calibrated inference can recover meaningful population dynamics that are not observable in static census products while avoiding the systematic distortions associated with uncalibrated or nationally normalized device counts.

Beyond the specific population product presented here, SAAC establishes a general methodological paradigm for population inference from opportunistic digital traces. The framework is modular, does not require prior knowledge of anchor population sizes, and can be implemented using data structures already present in many commercial mobility datasets. As such, it provides a practical foundation for evaluating existing population products, calibrating future high-frequency population estimates, and validating simulation-based mobility models.

As mobile sensing continues to expand in scope and scale, the need for principled calibration methods will become increasingly critical. By reframing population estimation as an anchor-constrained inference problem rather than a direct scaling exercise, SAAC offers a path toward more reliable, interpretable, and scientifically grounded representations of dynamic human presence. The resulting hourly population surfaces provide new datasets for research and applications in urban science, public health, environmental exposure assessment, and emergency management, where understanding when and where people are is as important as knowing where they live.

## 5 Methodology

### 5.1 Study area and data

#### 5.1.1 Study area

The study area is the United States (U.S.), comprising approximately 331 million residents distributed across about 220,000 Census block groups (CBGs). The spatial extent, population size, and heterogeneity of settlement patterns make this setting representative of large-scale applications of mobility-derived population estimation and comparable to prior national-scale studies. The U.S. also provides a diverse range of urban, suburban, and rural contexts in which observation coverage and device penetration vary substantially, posing a stringent test for calibration-based population inference.

#### 5.1.2 Mobility data

We use two complementary smartphone-based mobility datasets provided by Advan for the year 2022: Advan Neighborhood Patterns and Weekly Patterns (Table 1).

Neighborhood Patterns (Advan Research 2022b) provide stop-based observations of anonymized mobile devices aggregated at the CBG level. A stop is defined as a location event generated by a single tracked device that remains within a CBG for more than one minute. The dataset reports hourly stopped-device counts for each CBG over a calendar month, along with the inferred home CBGs of the observed devices and corresponding monthly counts. In addition, Advan reports the number of tracked devices associated with each home CBG, which enables estimation of device sampling rates. Neighborhood Patterns serve as the primary data source for modeling spatiotemporal population presence and flows among CBGs.

Weekly Patterns (Advan Research 2022c) provide visit-based observations at the point-of-interest (POI) level. A visit is defined as a recorded device ping within a POI polygon, regardless of dwell time. For each POI and calendar week, the dataset reports hourly visit counts and the number of distinct devices observed during that week. Although Advan uses the term "visitor," both stop-based and visit-based datasets fundamentally record device-generated events associated with distinct mobile devices. For conceptual consistency with the sampling model, we use the term *device* throughout this paper. Although the operational definitions of "stops" and "visits" differ across datasets, both represent discrete, device-generated location observations; therefore, we refer to them generically as *events* in the methodological development that follows.

The two datasets capture different aspects of the observation process: Neighborhood Patterns characterize spatially continuous stop-based presence at the CBG level, while Weekly Patterns provide dual observations of visits and distinct devices at selected POIs. Their joint use enables anchor-based calibration of observation scaling factors within the SAAC framework.

Table 1 summarizes the complementary structure of the two mobility datasets used in this study, highlighting how stop-based CBG observations provide continuous spatial coverage, while visit-based POI observations enable calibration of event-to-device relationships under stable-attendance conditions.

| Attribute | Neighborhood Patterns | Weekly Patterns |
|---|---|---|
| **Description** | Stop-based device events aggregated at the Census Block Group (CBG) level | Visit-based device events aggregated at the point-of-interest (POI) level |
| **Event definition** | A device-generated location event in which a tracked device remains within a CBG for more than one minute | A device-generated location event recorded within a POI polygon, regardless of dwell time |
| **Role in methodology** | Primary observation of hourly population presence and mobility flow distributions | Calibration of the event-to-device relationship under stable-attendance conditions |
| **Spatial unit of event observation** | CBG | POI |
| **Temporal resolution of event observation** | Hourly | Hourly |
| **Aggregation period for distinct devices** | Monthly | Weekly |
| **Example of hourly event counts** | Hourly stop counts: [10, 30, 20, 32, …] | Hourly visit counts: [21, 90, 32, 62, …] |
| **Example of distinct device origin distribution** | Monthly home-CBG counts: {$CBG_1$: 60, $CBG_2$: 63, …} | Weekly home-CBG counts: {$CBG_1$: 20, $CBG_2$: 33, …} |

### 5.1.3 Stable-attendance anchors

To identify stable-attendance anchors, we compiled a national inventory of U.S. high schools from high-schools.com, comprising approximately 30,000 public and private schools. High schools exhibit highly regular attendance during weekday school hours and are comparatively less influenced by discretionary mobility than many other public venues.

Students in grades 11 and 12 are typically aged 16–17, and smartphone ownership in this age group exceeds 95% (Faverio 2025). Moreover, despite restrictive school policies, most students actively use their devices on campus (Radesky et al. 2023). Advan's device panel includes users aged 16 and older (DeweyData, 2025), ensuring that high school attendance is well represented in the observed device population.

These characteristics make high schools well suited as stable-attendance anchors for examining event–device relationships under relatively homogeneous and temporally regular mobility conditions. High school POIs are used exclusively for calibration and diagnostic purposes within SAAC and are not included as target locations in the population estimation results.

### 5.1.4 Baseline population data

Residential population counts are obtained from the American Community Survey (ACS) 2019 5-year estimates, which align with the 2010 Census CBG boundaries used in Neighborhood Patterns. More recent ACS estimates rely on 2020 Census boundaries and are therefore not directly compatible with the mobility data without complex spatial reconciliation. The difference in total population between ACS 2019 and subsequent estimates is approximately 1.8%, which does not materially affect the inferred hourly population dynamics or the conclusions of this study. The ACS data provide the residential baseline population used in the population balance model underlying the SAAC framework.

## 5.2 CBG inbound population estimation

SAAC provides a scaling factor to estimate the latent devices $\hat{u}$ by event observations $e$. We estimate the CBG hourly inbound population $In_c^t$ as:

$$In_c^t = \hat{u}_c^t P_c^t = k_c^t e_c^t P_c^t \qquad (2)$$

where $\hat{u}_c^t$ denotes the estimated number of traced devices present in CBG $c$ at hour $t$, corrected for incomplete recording of stop events $\hat{u}_c^t = k_c^t e_c^t$, and $P_c^t$ denotes destination-specific device-to-person expansion factor at that time in CBG $c$.

$$P_c^t \approx \sum_{j=1}^{n} r_j^m P_j^m \qquad (3)$$

where $r_j^m$ denotes the proportion of observed devices originating from CBG $j$ in month $m$, and $P_j^m = N_j/D_j^m$ is the device-to-person expansion factors in the origin CBG, i.e., the reciprocal of the device sampling rate for CBG $j$ (Figure 5). $D_j^m$ refers to the tracked device count in CBG $j$, usually provided by data vendors. $N_j$ is the ACS population of CBG $j$.

Because the hourly CBG–level scaling factor $k_c^t$ is difficult to identify directly, we approximate it using a county–month aggregate scaling factor,

$$k_c^t \approx k_{cty}^m \qquad (4)$$

Under this approximation, the inbound population of Census block group $c$ at hour $t$ is reconstructed as:

$$In_c^t = k_{cty}^m e_c^t P_c^t \qquad (5)$$

To derive the county–month scaling factor $k_{cty}^m$, we aggregate weekly anchor-derived estimates within each month. Specifically, for all anchor school days in a given month, we sum the weekly distinct device counts and the corresponding typical school-hour event counts, and define the monthly scaling factor as the ratio of these two totals. This aggregation reduces week-to-week noise while preserving the stable event-to-unit relationship identified through SAAC. Additional technical details are provided in the Supplementary Information.

Intuitively, $P_j^m$ accounts for how many individuals are represented by a single observed device from origin CBG $j$, whereas $k_{cty}^m$ accounts for how many latent devices are associated with each recorded event in the observation process.

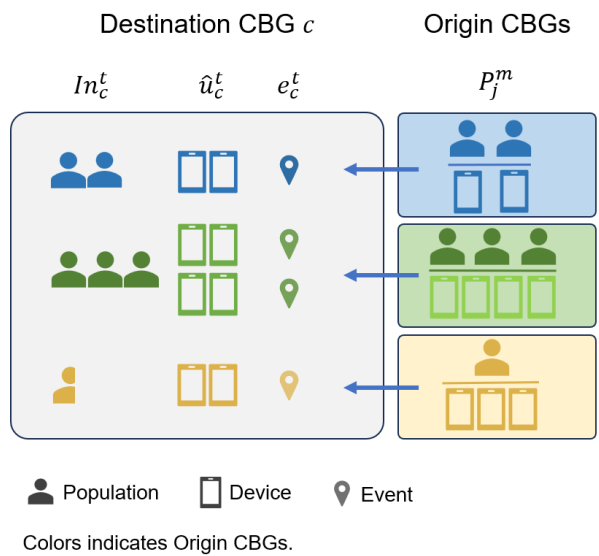

Figure 5 Conceptual illustration of estimating hourly inbound population presence at a destination CBG from mobility observations. The figure depicts the relationship between latent population presence, device-level observations, and event-based mobility signals for a destination CBG $c$ at hour $t$. Individuals (population) are partially represented by mobile devices, which generate observable events when visiting the destination. Observed event counts $e_c^t$ reflect an incomplete observation of the underlying event-generating *units*. Using anchor-based calibration, observed *events* are converted to an estimated number of distinct devices $\hat{u}_c^t$ via an observation model $\hat{u}_c^t = k_{cty}^m \cdot e_c^t$, where $k$ is the observation scaling factor inferred from stable-attendance anchors. Device origins are assigned to residential CBGs $j$, linking destination observations to origin-side residential population constraints $P_j^m$. Aggregating calibrated device counts across origins yields the estimated hourly inbound population presence $In_c^t$, expressed in population-equivalent *units* rather than raw device counts.

### 5.3 CBG outbound population estimation

Unlike inbound population presence, outbound mobility is generally not directly observable from opportunistic mobility data. However, mobility datasets typically contain the distribution of origin areas associated with observed mobility events. Assigning tracked devices to residential home areas (e.g., CBGs or geohash areas) is a common practice in mobility data construction, enabling linkage between device-based observations and resident-based demographic and sociological surveys. When device-generated events are observed at destination locations, each event can be interpreted as a movement from the device's inferred home CBG to the destination CBG. Aggregation of these origin–destination–linked events yields a distribution of device origins associated with each destination and observation period. This origin distribution provides the key auxiliary information required to reconstruct outbound population flows under incomplete and destination-biased observation.

Estimating the hourly outbound population at the CBG $c$, $Out_c^t$, is non-trivial because only partial summaries of *Origin–Destination–Time* flows are available (Figure 6). Conceptually, human mobility can be represented as a three-dimensional *Origin–Destination–Time* mobility cube, $M \in \mathbb{R}^{\tau \times n \times n}$, where $\tau$ denotes the number of hourly time steps in a month and $n$ denotes the number of CBGs. Each element $v_{jc}^t$ represents the number of individuals traveling from origin CBG $j$ to destination CBG $c$ at hour t, see Figure 6 (a).

If all cube elements were observed, the hourly outbound population of each CBG could be obtained by marginalizing the cube along the destination dimension, $Out_c^t = \sum_j v_{cj}^t$, which collapses the cube into a two-dimensional *Time-Orign* matrix of size $\mathbb{R}^{\tau \times n}$, i.e., blue face in Figure 6(b). Similarly, the hourly inbound population can be obtained by marginalizing the cube along the origin dimension, $In_c^t = \sum_j v_{jc}^t$, yielding the corresponding *Time–Destination* projection, i.e., Figure 6(c).

In practice, the full *Origin–Destination–Time* mobility cube is rarely available at an hourly resolution due to data volume and privacy constraints. Instead, opportunistic mobility datasets typically provide temporal aggregations along the time dimension, most commonly in the form of weekly or monthly *Origin–Destination* matrices. These aggregated *Origin–Destination* matrices correspond to a collapsed representation of the mobility cube, i.e., the purple face in Figure 6(a).

As a result, the hourly outbound population of CBG $c$, denoted as $Out_c^t$ (the *Time–Origin* matrix; blue face in Figure 6(b)), cannot be directly observed. It must be inferred jointly from the aggregated *Origin–Destination* information and the estimated hourly inbound population $In_c^t$ (the *Time–Destination matrix*, yellow face in Figure 6(c)). In other words, the *Time–Origin* matrix is reconstructed by jointly leveraging the *Time–Destination* and aggregated *Origin–Destination* marginals. This reconstruction problem is addressed using an iterative proportional fitting (IPF) procedure, which estimates latent hourly outbound flows by enforcing consistency between observed marginal totals and inferred hourly inbound population presence.

Specifically, we construct a two-dimensional matrix of size $\mathbb{R}^{\tau \times n}$ with hours as rows and origin CBGs as columns, corresponding to the *Time–Origin* marginal of the latent mobility cube. The row marginals are constrained to equal the total outbound population at each hour, which must equal the total inbound population across all destination CBGs at that hour, $\sum_c Out_c^t = \sum_c In_c^t$, where $In_c^t$ is estimated using the SAAC-calibrated inbound model described in Section 5.2. The column marginals are constrained to equal the total outbound population of each origin CBG over the calendar month, derived from Neighborhood Patterns.

Explicitly constructing and storing the full mobility cube is also an alternative, but it is computationally challenging at the national scale, as it would involve approximately 720×220,000×220,000≈10$^{13}$ latent elements, posing difficulties in both computation and data storage. To avoid explicit cube reconstruction, we used IPF.

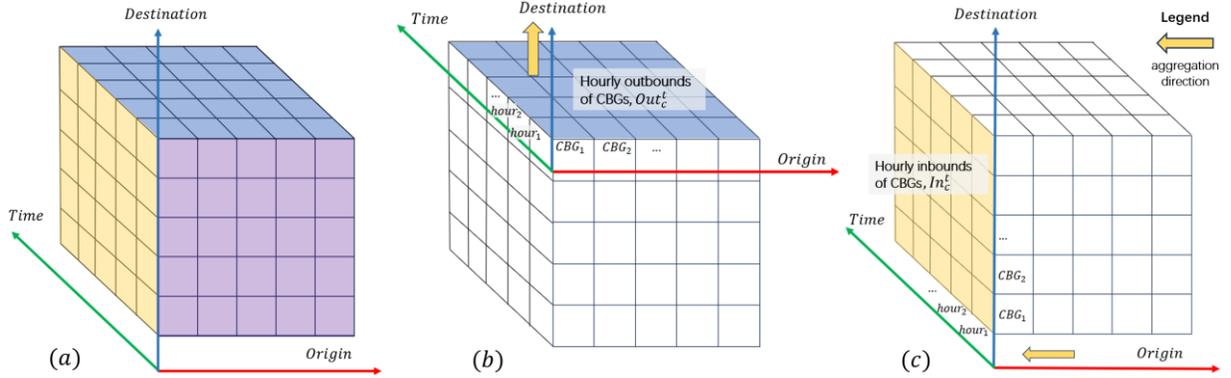

Figure 6 Conceptual illustration of the spatiotemporal mobility cube and its marginal aggregations. (a) A three-dimensional mobility cube indexed by time ($t$), origin $CBG_j$, and destination $CBG_c$. Each element represents the number of visitors traveling from origin $CBG_j$ to destination $CBG_c$ during hour $t$. (b) Aggregation of the mobility cube along the *Destination* dimension yields the hourly outbound population for each origin CBG, $Out_c^t$, forming a two-dimensional *Time–Origin* matrix (blue face). (c) Aggregation along the origin dimension yields the hourly inbound population for each destination CBG, $In_c^t$, forming a two-dimensional *Time–Destination* matrix (yellow face). Although the full mobility cube is not directly observable in opportunistic mobility data, its marginal aggregations are partially available. The hourly outbound population (blue face) is therefore reconstructed by jointly leveraging the inferred inbound population (yellow face) and the temporally aggregated *Origin–Destination* information (purple face), under consistency constraints.

## 5.4 Implementation

Explicitly materializing a full origin×destination×hour population matrix at national scale would be computationally prohibitive given the number of Census block groups and temporal steps involved. Accordingly, the proposed framework avoids storing or operating on a dense three-dimensional population cube. Stable-attendance anchor calibration is performed independently for each anchor and requires minimal computation. All raw inputs and intermediate results are instead maintained as two-dimensional, column-oriented tables, with spatial and temporal dimensions encoded as attributes rather than explicit axes. This design exploits the sparsity and skewed distribution of mobility observations and avoids combinatorial growth.

Inbound and outbound population components are inferred from aggregated marginal tables, and iterative proportional fitting (IPF) is applied at the marginal level rather than on a full origin–destination–time tensor, ensuring computational tractability. The implementation uses DuckDB as the primary analytical backend, coupled with a Python-based processing pipeline. DuckDB's columnar storage enables efficient filtering, aggregation, and joins across spatial and temporal dimensions, making the approach scalable to nationwide datasets covering approximately 220,000 Census block groups. Although the experiments were conducted on a machine with 160 GB of memory due to data volume, the implementation is intentionally designed to minimize memory usage and computational overhead by leveraging online analytical processing (OLAP) principles, thereby enhancing the reproducibility and replicability of the proposed method. Our code and reconstructed hourly population for all Census block groups in 2022 are available via the link in the "Data and code availability statement."

## 5.5 Evaluation

Direct validation of hourly population presence at fine spatial resolution remains challenging due to the scarcity of publicly available ground-truth data on dynamic populations. To contextualize and evaluate the plausibility of the SAAC-based reconstruction, we compiled several civic events with reported visitor counts and compared them against our estimated population spikes. In addition to these event-specific assessments, we conducted a national-scale comparison with the LandScan diurnal population products (daytime/nighttime). The Manhattan case study in Section 2.4 also serves as an illustrative benchmark. Collectively, these comparisons offer a perspective on the plausibility and consistency of the reconstructed results, rather than a formal accuracy assessment.

### 5.5.1 Comparison with reported visitor counts during local festivals

To provide face-validating evidence for the SAAC-calibrated population estimates, we examined hourly population dynamics for selected small towns that hosted well-known festivals in 2022. These towns typically have a residential population of around 1,000 to 2,000 people or fewer, but their annual events attract more than 10,000 visitors. Such sudden increases in population can be clearly observed in the reconstructed data. We intentionally selected one-day events instead of multi-day festivals to enable a more direct comparison between the estimated peak population and the officially reported attendance figures. Events without reliable or publicly reported visitor counts in 2022 were excluded to ensure the validity of the benchmark comparisons.

In several cases, we identified clear population spikes that coincided with the date of local events and found that the magnitudes of those spikes were reasonably consistent with publicly reported attendance numbers. For instance, the reconstructed population surges in Belzoni, MS (Catfish Festival) (World Catfish Festival 2022), Benson, NC (Mule Days) (Johnston County Visitors Bureau 2024), Caldwell, TX (Kolache Festival) (Burleson County Tribune 2022), and West Cape May, NJ (Christmas Parade) (Capemay.com 2022) closely matched the official or media-reported visitor estimates (Figure 7(a)-(d)). These examples suggest that the SAAC framework can detect and approximate population increases associated with planned local gatherings.

However, per our observation, not all known events produced detectable population spikes, and not all observed spikes could be linked to publicly documented events. In some cases, estimated surges were substantially different from reported figures. For example, in Far Hills, NJ (Far Hills Race Meeting) (Farhillsrace.org 2022), the reconstructed peak reached about 60,000, while the official report suggested around 30,000 attendees (Figure 7(e)). These discrepancies may arise from differences in measurement definitions, event duration, data limitations, or reporting inconsistencies. Moreover, some spikes could reflect data artifacts unrelated to real-world gatherings. This analysis highlights the potential of SAAC-based estimates to identify large-scale local events but also underscores the need for caution in interpreting individual spikes. A broader evaluation of accuracy, event detectability, and potential error sources will require more comprehensive benchmarking in future work.

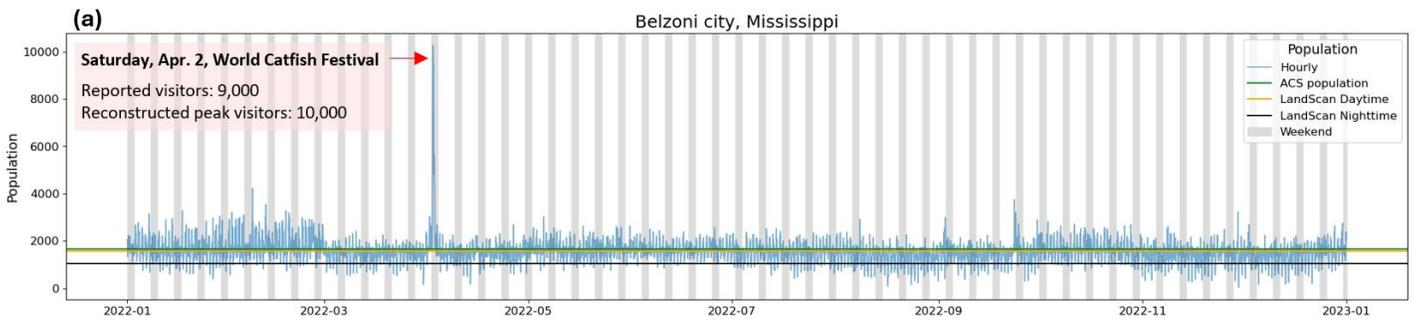
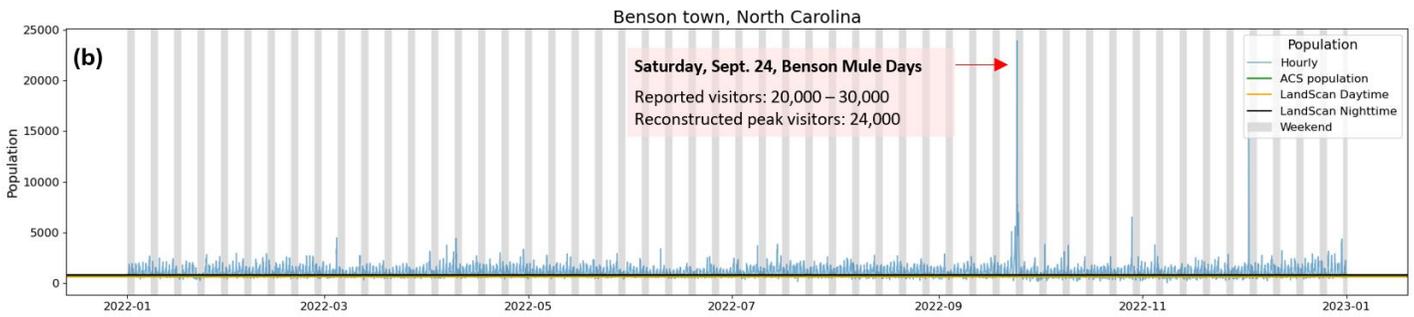
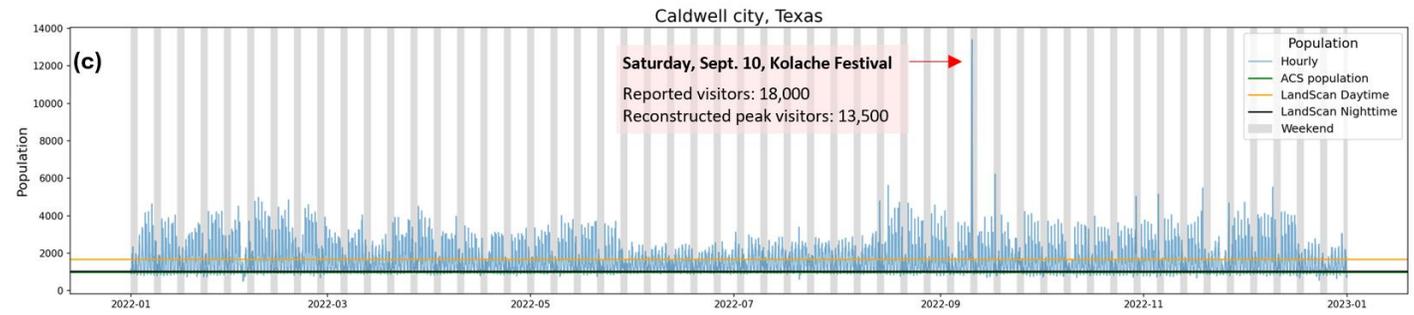
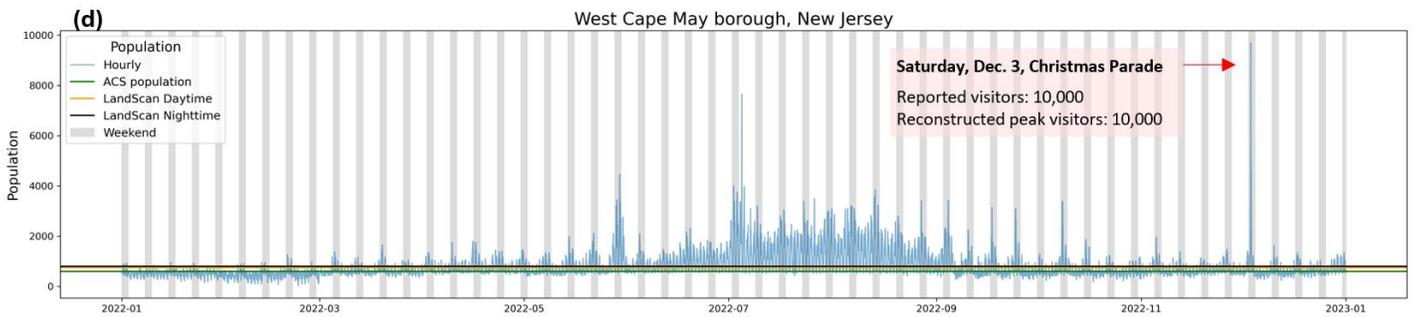
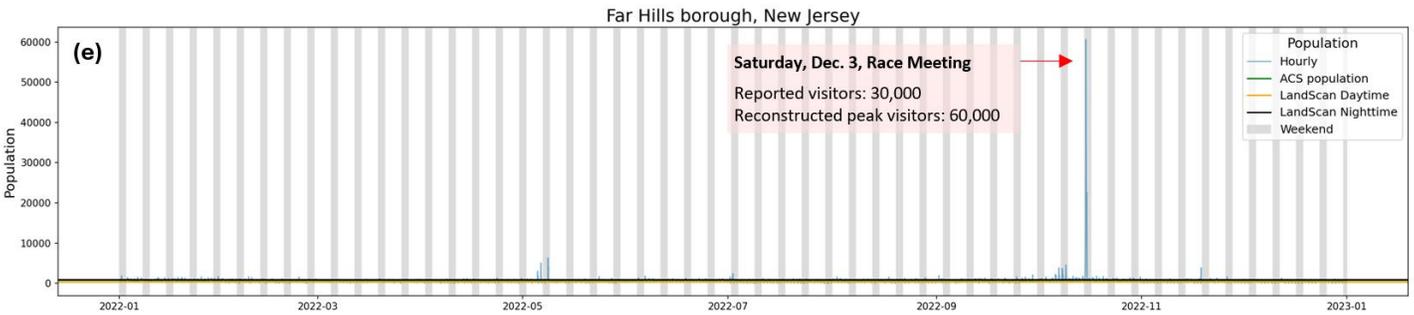

Figure 7 Reconstructed hourly population dynamics in five U.S. towns during known 2022 local festivals, based on SAAC-calibrated mobility data. Each panel shows a distinct one-day event: (a) Belzoni, MS (Catfish Festival), (b) Benson, NC (Mule Days), (c) Caldwell, TX (Kolache Festival), (d) West Cape May, NJ (Christmas Parade), and (e) Far Hills, NJ (Race Meeting). For each event, a clear population spike is observed, with reconstructed peak visitor estimates aligning closely with publicly reported attendance numbers in most cases. These examples demonstrate the SAAC framework's ability to detect large short-term population surges. However, not all events yield visible spikes, and some reconstructed peaks deviate from reported figures, highlighting both the promise and limitations of using mobility-based estimates for event detection and population quantification.

### 5.5.2 Comparison with LandScan gridded population product

We used the LandScan HD U.S. gridded population dataset (Weber et al., 2022) as an external reference to contextualize and assess the plausibility of the SAAC-based reconstruction nationwide, LandScan provides ambient population estimates that approximate the average population distribution over a 24-hour period by integrating land use and infrastructure information. Although LandScan does not resolve intra-day temporal variation and should not be interpreted as ground truth for hourly population presence, it represents one of the most widely used benchmarks for non-residential population distribution.

LandScan data are provided at approximately 100 m spatial resolution and are commonly used in population dynamics, exposure assessment, and disaster response research. Importantly, LandScan is not used as an input to the SAAC framework. Instead, it serves solely as an independent comparative reference to evaluate whether the magnitude and spatial structure of the reconstructed population are consistent with established ambient population products under daytime and nighttime conditions.

To facilitate comparison, we aggregate LandScan daytime and nighttime population estimates to CBG boundaries and compare them with corresponding SAAC-based estimates. Specifically, for each calendar month, we compute the average weekday noon population (12:00 PM local time) and compare it with the LandScan daytime population at the CBG level. Similarly, we compare the estimated midnight population (12:00 AM local time) with the LandScan nighttime population. Because LandScan is not a dynamic hourly dataset and does not represent a definitive ground truth, we report relative differences rather than errors.

The relative difference for each CBG is defined as:

$$diff_c^{rel} = \frac{\overline{Popu_c^{noon}} - Popu_c^{LandScan}}{Popu_c^{LandScan}} \qquad (6)$$

where:

$diff_c^{rel}$: relative difference of the CBG $c$.

$\overline{Popu_c^{noon}}$: average weekday noon (12: 00 PM, local time) population of the CBG $c$ in a calendar month.

$Popu_c^{LandScan}$: daytime population of the CBG $c$ in LandScan gridded population data (the latest release is 2021 when conducting this study).

Using this evaluation approach, the relative difference between the SAAC-based population estimates and LandScan averages 53.1% during daytime hours and 36.8% during nighttime hours across 2022 (Table 2). These differences reflect systematic conceptual distinctions between the two population constructs rather than estimation error. For example, LandScan nighttime population differs from the ACS 2019 residential population by approximately 20.6%, underscoring that substantial variation exists even among widely used population products.

Overall, the comparison indicates that the SAAC-based reconstruction produces population magnitudes that are broadly consistent with established ambient population datasets while additionally resolving intra-day temporal dynamics that are not captured by static products such as LandScan. The observed differences are therefore expected and reflect the complementary nature of dynamic mobility-derived population estimation versus static or quasi-static population representations.

Table 1 Difference between mobility-based population and LandScan gridded population product

| Month | 1 | 2 | 3 | 4 | 5 | 6 | 7 | 8 | 9 | 10 | 11 | 12 |
|---|---|---|---|---|---|---|---|---|---|---|---|---|
| Noon/daytime (%) | 52.7 | 51.1 | 49.3 | 50.2 | 52.3 | 53.6 | 58.7 | 54.6 | 51.5 | 52.1 | 54.6 | 59.1 |
| Midnight/nighttime (%) | 37.4 | 36.6 | 35.7 | 35.9 | 35.9 | 36.0 | 38.1 | 37.3 | 36.8 | 36.6 | 37.7 | 37.6 |

**Data and code availability statement**

Advan Neighborhoods Patterns and Weekly Patterns can be subscribed from Dewey Inc. (deweydata.io). Census data can be downloaded from U.S. Census Bureau (census.gov). The high school list can be subscribed to from https://high-schools.com.

The data processing code is available at https://github.com/gladcolor/hourly_population. The reconstructed hourly population data for all U.S. neighborhoods (about 220,000 Census block groups) in 2022 are available at:
https://huggingface.co/datasets/gladcolor/hourly_population_US.

Supplementary Information

# Nationwide Hourly Population Estimating at the Neighborhood Scale in the United States Using Stable-Attendance Anchor Calibration

## List of Supplementary Items

S1. Notation Used in the Stable-Attendance Anchor Calibration (SAAC) Framework

S2. Identification of School Days

S3. Identification of Non-School-Hour Gatherings

S4. Identification of Valid Anchor Weeks

S5. Composition of County–Month Observation Scaling Factors

# S1. Notation Used in the Stable-Attendance Anchor Calibration (SAAC) Framework

Table S1 Notation used in the SAAC framework

| Symbol | Description |
|---|---|
| $c$ | Index of destination Census Block Group (CBG) |
| $j$ | Index of origin Census Block Group (CBG) |
| $t$ | Hour index within a calendar month |
| $\tau$ | Total number of hourly time steps in the month |
| $n$ | Total number of Census Block Groups |
| $e$ | Typical observed event count in a stable-attence window |
| $e_c^t$ | Observed device-event count in a stable-attence window in CBG $c$ at hour $t$ |
| $u$ | Count of event-generating units present during a stable attendance window |
| $\hat{u}$ | Stabilized cumulative count of distinct observed event-generating units |
| $\hat{u}_c^t$ | Estimated count of traced devices present in CBG $c$ at hour $t$, corrected for incomplete recording of event, $\hat{u}_c^t = k_c^t e_c^t$ |
| $N_c$ | Residential population of CBG $c$ from the American Community Survey (ACS) |
| $p_c^t$ | Population present in CBG $c$ at hour $t$ |
| $In_c^t$ | Inbound population entering CBG $c$ at hour $t$ |
| $Out_c^t$ | Outbound population leaving CBG $c$ at hour $t$ |
| $C_c^t$ | Observed stopped-device count in CBG $c$ at hour $t$ |
| $D_j^m$ | Number of tracked devices residing in origin CBG $j$ in month $m$ |
| $P_j^m$ | Device-to-person expansion factor for origin CBG $j$ in month $m$ |
| $P_c^t$ | Destination-specific device-to-person expansion factor for destination CBG $c$ at hour $t$ |
| $k$ | Observation scaling factor (OSF) correcting for incomplete stop or visit recording |
| $k_c^t$ | Observation scaling factor for destination CBG $c$ at hour $t$ |
| $k_{cty}^m$ | County–month observation scaling factor estimated using stable-attendance anchors |
| $r_j^m$ | Proportion of stopped devices in a destination CBG originating from CBG $j$ during month $m$ |
| $v_{jc}^t$ | Latent visitor flow from origin CBG $j$ to destination CBG $c$ at hour $t$ |
| **M** | Spatiotemporal mobility cube indexed by time, origin, and destination |
| **M**$(t, j, c)$ | Element of the mobility cube representing visitor flow $v_{jc}^t$ |
| OSF | Observation scaling factor correcting event-level under-recording |
| SAAC | Stable-Attendance Anchor Calibration framework |

## S2. Identification of School Days

To operationalize the Stable-Attendance Anchor Calibration (SAAC) framework, we first identify valid school days for each high school point of interest (POI) using hourly event time series from Weekly Patterns. The objective is to distinguish routine instructional days from holidays, half days, and non-instructional days based solely on mobility-derived temporal signatures, without relying on external school calendars.

For each school–week record, weekly visit profiles were expanded into hourly event time series using the reported visits_by_each_hour field. To reduce high-frequency noise and transient reporting fluctuations, hourly event counts were smoothed using a Savitzky–Golay filter with a 13-hour window and a third-order polynomial. Smoothed values were constrained to be non-negative.

School days were identified using peak-based detection of stable daytime attendance. Local peaks in the smoothed hourly event series were detected under the following criteria: a minimum peak height of 15 events, minimum prominence of 8 events, minimum inter-peak distance of 12 hours, and minimum peak width of 3 hours. These thresholds were selected empirically to capture sustained school-hour attendance while excluding minor fluctuations.

Because attendance levels vary systematically across the academic year, semester-specific attendance baselines were estimated for spring (February–May) and fall (September–November). For each semester, weekday daytime peak event counts were extracted, and the median of these values was used as the semester-specific baseline. Dates before July 1 were evaluated against the spring baseline, while dates on or after July 1 were evaluated against the fall baseline.

A calendar date was classified as a school day if it satisfied both of the following conditions:

1. at least one daytime attendance peak was detected on that date; and
2. the peak event count exceeded one-half of the semester-specific baseline.

All weekend dates (Saturday and Sunday) were excluded from school-day classification regardless of event magnitude. This procedure retains half days with substantial instructional attendance while excluding holidays and non-instructional days.

## S3. Filtering of Non-School-Hour Gatherings

Even during valid school weeks, high schools may host **non-instructional gatherings**, such as sporting games, concerts, or community activities, that generate elevated mobility observations outside instructional hours. To prevent these gatherings from biasing the estimation of typical school-hour event levels, non-school-hour gatherings were explicitly identified and flagged.

After-school hours were defined as 17:00–21:59 local time on weekdays. For each identified school day, total after-school event counts were aggregated and compared against semester-specific medians, computed separately for spring and fall semesters. An after-school gathering was flagged if total after-school event counts exceeded twice the corresponding semester median. To avoid spurious detection driven by low-volume noise, after-school peaks with a maximum hourly event count below 10 were ignored.

Weekend daytime gatherings were analyzed separately. Weekend daytime hours were defined as 07:00–16:59 local time on Saturdays and Sundays. Total weekend daytime event counts were aggregated by date, and semester-specific medians were computed. A weekend daytime gathering was flagged if total event counts exceeded twice the semester median and the maximum hourly event count exceeded 20.

For each detected non-school-hour gathering, the hour with the maximum event count was recorded as the gathering peak. Dates associated with such gatherings were flagged to support subsequent week-level screening.

## S4. Identification of Valid Anchor Weeks

Observation scaling factors were estimated only from calendar weeks exhibiting stable instructional attendance with minimal contamination from **non-school-hour gatherings**. To identify such anchor weeks, daily and weekly summaries were constructed and evaluated against a set of exclusion criteria.

For each calendar week, the following quantities were computed:
- number of identified school days;
- total school-hour event counts during instructional hours (07:00–16:59 on school days);
- total after-school and weekend event counts;
- presence of detected evening or weekend gatherings;
- typical hourly event count, defined as the 80th percentile of weekday noon-hour (11:00) event counts within the week;
- weekly distinct observed units (devices) reported by Weekly Patterns.

A week was considered a valid anchor week if it satisfied all of the following conditions:
1. at least four identified school days were present;
2. the sum of the two largest non-school-hour gathering peak event counts did not exceed the typical hourly event count;
3. the ratio of non-school-hour gathering events (after-school plus weekend) to school-hour events did not exceed 0.2;
4. the implied visit-based observation scaling factor did not exceed 7.

These criteria ensure that retained weeks represent routine instructional activity with limited non-instructional gatherings and stable event–unit relationships, consistent with the SAAC stable-attendance assumption.

**S5. Composition of County–Month Observation Scaling Factors**

The Stable-Attendance Anchor Calibration framework estimates county–month observation scaling factors (OSFs) using school-day observations during periods with stable attendance. However, not all counties and months contain sufficient clean weeks with valid school-day observations. This occurs most frequently in small counties, rural areas, or months associated with school breaks and transitional periods. To ensure complete spatiotemporal coverage of OSFs, we applied a rule-based temporal and spatial completion procedure.

First, county–month OSFs were estimated directly from high school observations in months with sufficient clean weeks. For months lacking reliable school-day observations, OSFs were temporally completed using adjacent months according to the academic calendar structure:

1. January OSF was set equal to the February OSF.
2. June OSF was set equal to the May OSF.
3. August OSF was set equal to the September OSF.
4. July OSF was set as the arithmetic mean of the May and September OSFs.

These rules reflect the absence of stable instructional attendance during school breaks and ensure smooth seasonal continuity of OSFs while remaining anchored to empirically observed periods.

Second, for counties where county–month OSFs could not be reliably estimated due to insufficient school-day observations even after temporal completion, a spatial fallback strategy was applied. Specifically, state-wide OSFs were computed by aggregating raw device counts and typical hourly visits across all counties within the same state for each month. Counties lacking sufficient local observations in a given month were assigned the corresponding state-level OSF.

This hierarchical calibration strategy ensures full spatial coverage while preserving local estimates whenever supported by data. It also avoids extrapolation across states and maintains consistency with the institutional and behavioral context of stable-attendance anchors.